\documentstyle[aps,preprint]{revtex}

\draft

\begin{document}
%\columnsep -.375in
%\twocolumn[

\begin{title}
 {Effect of conduction electron interactions on Anderson impurities}
\end{title}

\author{Y. M. Li \cite{Li} }

\address{Max-Planck-Institut f\"{u}r Physik komplexer Systeme,
     Bayreuther Str.\ 40, Haus 16, 01187 Dresden, Germany}
\maketitle

\begin{abstract}
The effect of conduction electron interactions for an Anderson impurity
is investigated in one dimension using a scaling approach.
The flow diagrams are obtained by solving  the renormalization group
equations  numerically. It is found that the Anderson impurity
case is different from its counterpart --
the Kondo impurity case even in the local moment region.
The Kondo temperature for an Anderson impurity shows nonmonotonous behavior,
increasing for weak interactions but decreasing for strong interactions.
The implication of the study to other related impurity models is also
discussed.

\end{abstract}

\bigskip
\bigskip


%\newpage
%\narrowtext

Recently there has been much interest in
magnetic impuirties interacting with a one-dimensional (1D)
correlated fermion system \cite{Lee,FN,preprint,Karen,FG}.
On the one hand, the progress in the nanofabrication technology
could make the question  accessible experimentally. On the other hand,
it would shed some light on the  impurity  as well as
lattice systems  in $D>1$ in the presence of conduction electron correlations.
It has been found that localized electrons interact with
strongly correlated  conduction electrons
in many materials, particularly
in high-$T_c$ superconductors and the
heavy-fermion-like compound ${\rm Nd}_{1.8}
{\rm Ce}_{0.2} {\rm Cu} {\rm O}_4$ \cite{Nd}.

It is known that
electrons in 1D systems are in a Luttinger liquid state \cite{Haldane1}.
A Kondo impurity in a Luttinger liquid was studied recently by
Lee and Toner \cite{Lee}, and Furusaki and Nagaosa  \cite{FN}.
They found that the Kondo temperature $T_{\rm K}$
has an algebraic dependence on the Kondo
coupling rather than the exponential dependence of the usual Kondo impurity,
and  always rises as the strength of
conduction electron interactions increases.  In effect, there is no
competition between electron correlations and the Kondo effect.

The usual Anderson impurity  Hamiltonian with  a free conduction bath
can be transformed into the Kondo Hamiltonian with an irrelevant potential
scattering \cite{Wilson} using the Schrieffer-Wolff transformation
in the local moment regime.
However,  when the Schrieffer-Wolff transformation
is applied in the presence of conduction electron interactions,
the effective Kondo coupling strongly depends on the
interactions. Moreover, the impurity spin will interact
not only with conduction electron spins at the impurity site but also with
spins at neighboring sites \cite{Tom}.
The potential scattering is also relevant in a Luttinger
liquid \cite{KF,FG}.  One would expect that
an Anderson impurity  behave differently from  a Kondo impurity
when conduction interactions are included even in the local moment region.

In this paper, we mainly report  results for an Anderson impurity
in a Luttinger liquid.  Using the scaling approach of
Anderson-Yuval-Hamanna (AYH)
\cite{AYH} and Cardy \cite{Cardy}, we obtain the renormalization group (RG)
equations, which  are solved exactly by numerical methods.
We find that there is a strong interplay between the Kondo effect and the
electron interactions for the Anderson impurity,
unlike the Kondo impurity case.
The Kondo temperature is enhanced for weak electron interactions but
reduced for  the strong interacting case.
The zeroth-order approximation  which is widely used in the literature
for solving the RG equations is checked against our numerical results.  It
suggests that the zeroth-order approximation could be very misleading.

The Anderson impurity in a Luttinger liquid is described by the Hamiltonian
\begin{eqnarray}
\label{H}
H & = & H_{\rm L}+H_f+H_{c-f},    \nonumber \\
H_{\rm L} & =  & v_F \sum_{k,\sigma} k
           ( c^{\dag}_{k,1,\sigma} c_{k,1,\sigma}
            -c^{\dag}_{k,2,\sigma} c_{k,2,\sigma} ) \nonumber \\
 && + \frac{g_2}{N} \sum_{k_1,k_2,p,\sigma_1,\sigma_2}
      c^{\dagger}_{k_1,1,\sigma_1}   c^{\dagger}_{k_2,2,\sigma_2}
   	 c_{k_2+p,2,\sigma_2}  c_{k_1-p,1,\sigma}, \nonumber \\
H_f &=& E_f^0 \sum_\sigma n_{f\sigma} +
             Un_{f\uparrow} n_{f\downarrow} , \\
H_{c-f} &=&  \frac{t}{\sqrt{N}}\sum_{k,i,\sigma}
           (c^{\dag}_{k,i,\sigma}  f_{\sigma}+ H.c.), \nonumber
\end{eqnarray}
Where $c_{k,1,\sigma}$ ($c_{k,2,\sigma}$) is the annihilation operator of
right-moving (left-moving) electrons with spin $\sigma$ and momentum $k$,
$f_\sigma$ is the annihilation operator for localized electrons, $n_{f\sigma}
=f_\sigma^\dagger f_\sigma$, and $N$ is the number of lattice sites.
$H_{\rm L}$ is  the  so-called
Tomonaga-Luttinger Hamiltonian, where  the $g_2$ term represents
forward scattering. $H_{\rm L}$ generally describes  1D fermion
systems away from half-filling and with repulsive interactions, for which
umklapp and backward scattering can be ignored.
$H_f$ and $H_{c-f}$ are the local and mixing terms, respectively.
In the case of $g_2=0$, the total Hamiltonian
$H$ reduces to the usual Anderson Hamiltonian.

We will  derive the partition function
using bosonization technique \cite{boson} and
then  obtain the renormalization group equations
applying the scaling approach of Anderson-Yuval-Hamann \cite{AYH}
and Cardy \cite{Cardy}.
The partition function  of the system  is $Z=\int {\cal D}c {\cal D} f
\exp \left[ -S_0-\int_0^\beta d \tau H_{c-f} (\tau) \right] $, where
$S_0$ is the action corresponding to $H_{\rm L} +H_f$.
Paralleling the previous studies on impurity problems \cite{Haldane,SK},
the strategy for finding $Z$ is the following. First, $Z$ is
expanded  and  is written  in terms of
histories of the impurity.
In this paper we will take the on-site Coulomb repulsion $U \to \infty$.
The history for the $n$-th term is  thus a sequence of transitions,
taking place at the imaginary time $0<\tau_1<...<\tau_n<\beta$,
between the three local $f$ states $\mid\alpha \rangle$ with
$\alpha=0$ (i.e., the $f^0$ configuration),  $\sigma$ (here
$\sigma=\pm1$ stands for the $f^1$ configuration with spin-up and
spin-down, respectively).
The degrees of freedom of conduction electrons left in $Z$ are then treated
using bosonization.  This gives
\begin{equation}
 Z = Z_0 \sum_{n=0}^{\infty}
\sum_{\alpha,\alpha_1,..,\alpha_n } \sum_{l_1..l_n}
\int_{0}^{\beta}\frac{d\tau_{n}}{\xi}
\int_0^{\tau_n-\xi}  \frac{d\tau_{n-1}}{\xi}...\int_0^{\tau_2-\xi}
\frac{d\tau_1}{\xi}
\exp[-S_{ \{ \alpha l \} } (\tau_1 ... \tau_n)] ,
\label{Z}
\end{equation}
where $Z_0$ is the partition function for $H_{\rm L}+H_f$, and
the cut-off is $\xi= 1/W$
($W$ is the conduction electron bandwidth).  Furthermore,
$l_i$ in the summation  takes 1 and 2, labelling
the left and right movers, and
\begin{equation}
S_{ \{ \alpha  l \} }  (\tau_1 ... \tau_n)
= - \sum_{i<j}(-1)^{i+j} K^{l_i l_j}_{ij}
 {\rm ln} \frac{\tau_j-\tau_i}{\xi} -n \ln y_{\alpha_i,\alpha_{i+1}}
  + \sum_i h_{\alpha_{i+1}} \frac{\tau_{i+1}-\tau_i}{\xi}.
\label{S}
\end{equation}
The fugacity $y_{\alpha_i,\alpha_{i+1}}$ in (\ref{S}) is the amplitude
associated with a transition from $\mid\alpha_i>$ to
$\mid\alpha_{i+1}>$.   $y_{0 \sigma}=y_{\sigma 0}=
\sqrt{\Delta \xi/\pi}$, here $\Delta$ is
the bare hybridization strength, defined by $\pi \rho t^2$
($\rho$ is the bare density of states of the bath at Fermi level) as usual.
The effective ``magnetic
field'' $h_{\alpha_i}$ reflects the differences of the local state energies:
$h_0=-2 E_f /3$, $h_{\sigma}=E_f /3$.
$K^{l_i l_j}_{ij}$ corresponding to the scaling dimension
is given by
\begin{equation}
K^{l_i l_j}_{ij}=\frac{1}{2}  [ (1+2 \sinh^2\phi)\delta_{l_i l_j}
        + \sinh 2\phi (1-\delta_{l_i l_j})
     +m_i m_j \delta_{l_i l_j} ],
\end{equation}
where $\phi=(1/2) \tanh^{-1} (-g_2/\pi v_F)$ and
$m_i=\alpha_i+\alpha_{i+1}$. The long-range logarithmic interaction
in (\ref{Z}) arises
from the reaction of the conduction electron bath towards the transition
between the local states. So  $K^{l_i l_j}_{ij}$ describes
the reaction strength with respect to the local disturbance.

Using the partition function (\ref{Z}) and the scaling approach
\cite{AYH,Cardy},  we obtain the following RG equations
\begin{eqnarray}
 \frac{d{\Delta}}{dln\xi} &=&-\gamma{\Delta}, \label{RG1}\\
 \frac{dE_f}{dln\xi}&=&\frac{\Delta}{\pi}
     (2e^{-E_f\xi}-e^{E_f\xi}), \label{RG2}\\
 \frac{d\gamma}{dln\xi} &=& -4 (\gamma+1)\frac{\Delta\xi}{\pi}
        (2e^{-E_f\xi}+e^{E_f\xi}), \label{RG3}
\end{eqnarray}
where
\begin{equation}
\gamma= \frac{1}{2} \left[
        \frac{1}{\sqrt{ 1- [g_2/(\pi v_F)]^2 } }-1 \right].
\label{gamma}
\end{equation}
The parameter $\gamma$ is just the exponent of the  distribution
function, $n_\sigma (k)$, of conduction electrons at the Fermi surface
in the absence of the impurity, i.e.,
$n_\sigma (k) \sim |k-k_F|^{2\gamma}$.
It is noted that when the above RG equations
are derived,  only  the hybridyzation term $H_{c-f}$ is treated as
perturbation,
while the conduction electron interactions and the ``external field'',
i.e., the energy level of local electrons  are dealt with nonperturbatively.
Eqs.\ (\ref{RG1})-(\ref{RG3}) are complicated and
will be solved numerically.

The above RG equations  have been discussed
using the zeroth-order approximation in the literature.
In the zeroth-order approximation, the parameter $\gamma$ is not renormalized,
i.e., $d\gamma / d {\rm ln}\xi \approx 0$ since $\Delta\xi \ll 1$.
For $E_f \xi \ll 1$, Eq.\ (\ref{RG2}) is reduced to
\begin{equation}
 \frac{dE_f}{dln\xi}\approx\frac{\Delta}{\pi}.
\label{RG4}
\end{equation}
For $g_2=0$ (or $\gamma=0$), Eqs. (\ref{RG1}) and (\ref{RG4}) reproduce
Haldane's scaling equations for the usual Anderson model \cite{Haldane}.
For a constant $\gamma$, integrating (\ref{RG1}) and (\ref{RG4}),
we obtain the flowing resonance width and impurity level
\begin{eqnarray}
\Delta &=& \Delta_0 \left( \frac{W}{W_0} \right)^\gamma,
\nonumber \\
E_f &=&E_{f}^0-\frac{\Delta_0}{\pi \gamma}
     \left[ (\frac{W}{W_0})^\gamma -1 \right],
\label{traj}
\end{eqnarray}
where $\Delta_0$, $E_f^0$  and $W_0$ are initial (bare) values.
If the Kondo temperature $T_{\rm K}$  is estimated
from $W\approx -E_f\approx  \alpha T_{\rm K}$
(here   $\alpha$
is an universal number characteristic of the crossover) \cite{Haldane},
using (\ref{traj}) one can  obtain
\begin{equation}
T_{\rm K} \sim W_0 (1-\frac{\gamma}{J \rho})^{1/\gamma},
\label{Tk}
\end{equation}
where $J =\Delta_0/(\pi \mid E_f^0 \mid \rho)$.
When $\gamma \geq J\rho$, $T_{\rm K}$ becomes zero.

Now let us look at numerical solutions of the RG equations
(\ref{RG1}) - (\ref{RG3}).  The parameter
$\gamma$ as a function of the bandwidth
is shown in Fig.\ 1(a) for various initial values $\gamma_0$
(curves  labeled a, b and c  correspond to
$\gamma_0 =0.1, 0.5, 1.0$, respectively).
It is seen that $\gamma$ decreases fast as the
bandwidth reduces.  The corresponding resonance width $\Delta$ is
shown in Fig.\ 1(b),
compared with the result of the zeroth-order approximation (dashed curves).
Since the perturbative
scaling approach is valid in the region $\Delta \ll W$,
the straight dotted line   $\Delta=W$ marks the rough boundary beyond
which the renormalization is incorrect quantitatively. It is noted that
even in the region of $W>>\Delta$ where the renormalization
is valid, there is  a significant difference
between exact results  and those of the
zeroth-order approximation. The corresponding impurity level $E_F$ is shown
in Fig.\ 1(c).  It is different from the result of
the zeroth-order approximation even
for a large bandwidth $W \sim W_0$.  It is noted that the zeroth-order
approximation gives wrong flows of $E_F$ for small $W$ in the case of
large $\gamma$.

 One important observation on Fig.\ 1(a) and 1(b)
is that  at the
points  where $\gamma$ flows  to zero,  the corresponding
resonance width $\Delta$ is still much less than the bandwidth.
This means that our renormalization process is still valid  at $\gamma=0$.
One of the advantages of the scaling approach is that  the
systems with different parameters on the same flow line in the flow
diagrams are indicated to have
the same universal behavior.
Although this approach does not solve the model, one
can know  information  of other systems  if one of the systems on the
flow line is  soluble.
The $\gamma=0$ system is the usual Anderson impurity problem \cite{note},
which is the well-known exactly soluble case \cite{Wilson,BA}.
The Kondo temperature for the
usual Anderson impurity is $T_{\rm K} = W \sqrt{J \rho} \exp({-1/J\rho})$.
We will use this expression to estimate the Kondo temperature for our
system.  It is noted that all the parameters in the expression should be
replaced by the values of the correponding parameters at $\gamma=0$, not the
initial values.

We obtain the Kondo temperature shown in
Fig.\ 2 for  two  choices of initail values of parameters (solid curves).
$T_{\rm K}$ increases for small $\gamma_0$ but decreases for large $\gamma_0$.
This is consistent with  small cluster calculations \cite{Karen}.
Basically the conduction electron
interactions have two  effects
on the Kondo temperature. On the one hand,
more conduction electrons could participate in the screening on the
impurity spin in the interacting case
than in the free conduction electron case where
only the electrons near the Fermi surface can contribute to
this screening process,
so that the Kondo temperature is expected to be enhanced by the
conduction electron interactions.
On the other, the effective Kondo coupling decreases because
the second-order hopping processes contributing to the
spin-flip exchange, i.e., the Kondo scattering, in the Schrieffer-Wolff
transformation are unfavorable due to
the interactions \cite{Tom}. The competition of these two effects
results in the nonmonotonous behavior of the Kondo temperature.
For the Kondo impurity model, in contrast,  the Kondo coupling
is assumed to be constant  so that the Kondo temperature is
always enhanced by the conduction electron interactions.

The  correponding Kondo temperatures in the zeroth-order approximation are
also shown in  Fig.\ 2 for comparison (the dashed curves).
It is seen that the zeroth-order approximation gives a sharp decrease for
$T_{\rm K}$ as $\gamma_0$ increases and predicts a transition at
$\gamma_0 = J \rho$.  This  artificial transition  originates
from the incorrect flows of
$E_F$, shown in Fig.\ 1(c), where the dashed curves do not tend to zero
for small $W/W_0$ in the case of large $\gamma_0$.

Incidentally, we would like to give a remark on
the usual Anderson model
with  the screening channels.  This is a model
which has attracted much attention recently
since it could show non-Fermi liquid behavior \cite{c-f,ZY,Yu,PRB}.
The  RG equations of this model have the same form
as  those given in  (\ref{RG1})-(\ref{RG3}).
The parameter $\gamma$ included is given
by the phase shifts of the conduction
electrons scattering from the impurity.
So the above discussions on the RG equations
are also applicable to this case.
The Kondo temperature should have the
same behavior as that shown in Fig.\ 2.  The renormalization of
$\gamma$ has not been taken into account  in discussing
physical properties of the system in the literature.
The present study strongly suggests that
it is important to include  this renormalization
for a complete understanding of this system.

In conclusion, we have found that there is a basic difference
between an Anderson and a Kondo
impurity in the presence of conduction electron interactions
even in the local moment regime.
The conduction  electron interactions  enhance the Kondo temperature for the
weak case, but will compete with the Kondo effect for the strong  case.
Although the present study is confined in one dimension,
we expect that this effect  exist for
dimensions higher than one.  We have demonstrated  that the
renormalization of the  parameter related to
interactions  is essential for a consistent scaling theory.
This is also true for other  related impurity models.

The author is grateful to  N. d'Ambrumenil,
K. A. Hallberg,  T. Schork,
Z. B. Su,  E. Tosatti,   L. Yu,
and Y. Yu for useful discussions,  especially to P. Fulde
for stimulating discussions,  constant encouragement and
warm hospitality extended to him during  his stay in Dresden.

%\newpage

\begin{description}

\item{Fig.\ 1}

The flow diagrams: (a)  the parameter $\gamma$ related
to the interactions; (b) the
resonance width $\Delta$ and (c) the impurity level $E_F$,
as functions of the bandwidth $W$ for various initial values of $\gamma_0$.
Curves labeled  $a, b, c$ are for $\gamma_0$ $=0.1, 0.5, 1.0$, respectively.
The initial values for $E_F^0$ and $\Delta_0$ are  chosen to be $-0.15$ and
$0.2 \pi |E_F^0|$, respectively.
The dashed lines in (b) and (c) are the corresponding quantities
in the zeroth-order approximation [given in (\ref{traj})].
The dotted line in (b) is $\Delta=W$.  All energies are in unit of $W_0$.

\item{Fig.\ 2}

The Kondo temperature  $T_{\rm K}$
as a function of  $\gamma_0$ for two choices of initial
parameters.  Curve a:  $E_F^0=-0.15$, $\Delta_0=0.2 \pi |E_F^0|$;
Curve b:  $E_F^0=-0.1$, $\Delta_0=0.15 \pi |E_F^0|$. The
dashed curves are the corresponding results of the zeroth-order approximation
[given in (\ref{Tk})].

\end{description}


\begin{references}
\bibitem{note} For $\gamma=0$, the impurity spin is coupled
       with free left and right moving electrons, which is reminiscent of the
        two-channel Kondo problem. However, the fixed point would be unstable
       for the asymmetry of the Kondo couplings of the impurity spin
      with spins at neighboring sites and
    the backward Kondo scattering \cite{FN}.
\end{references}
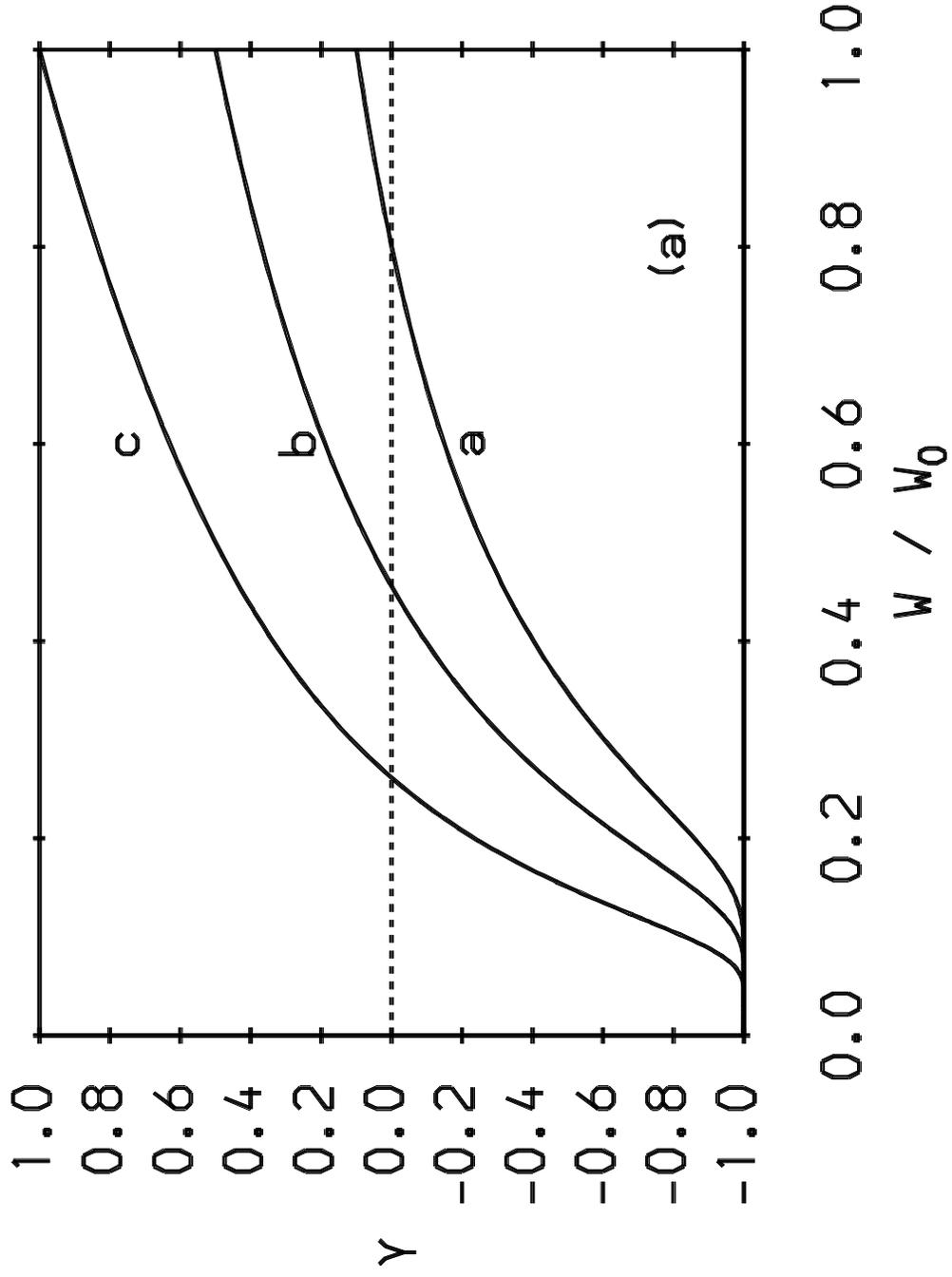
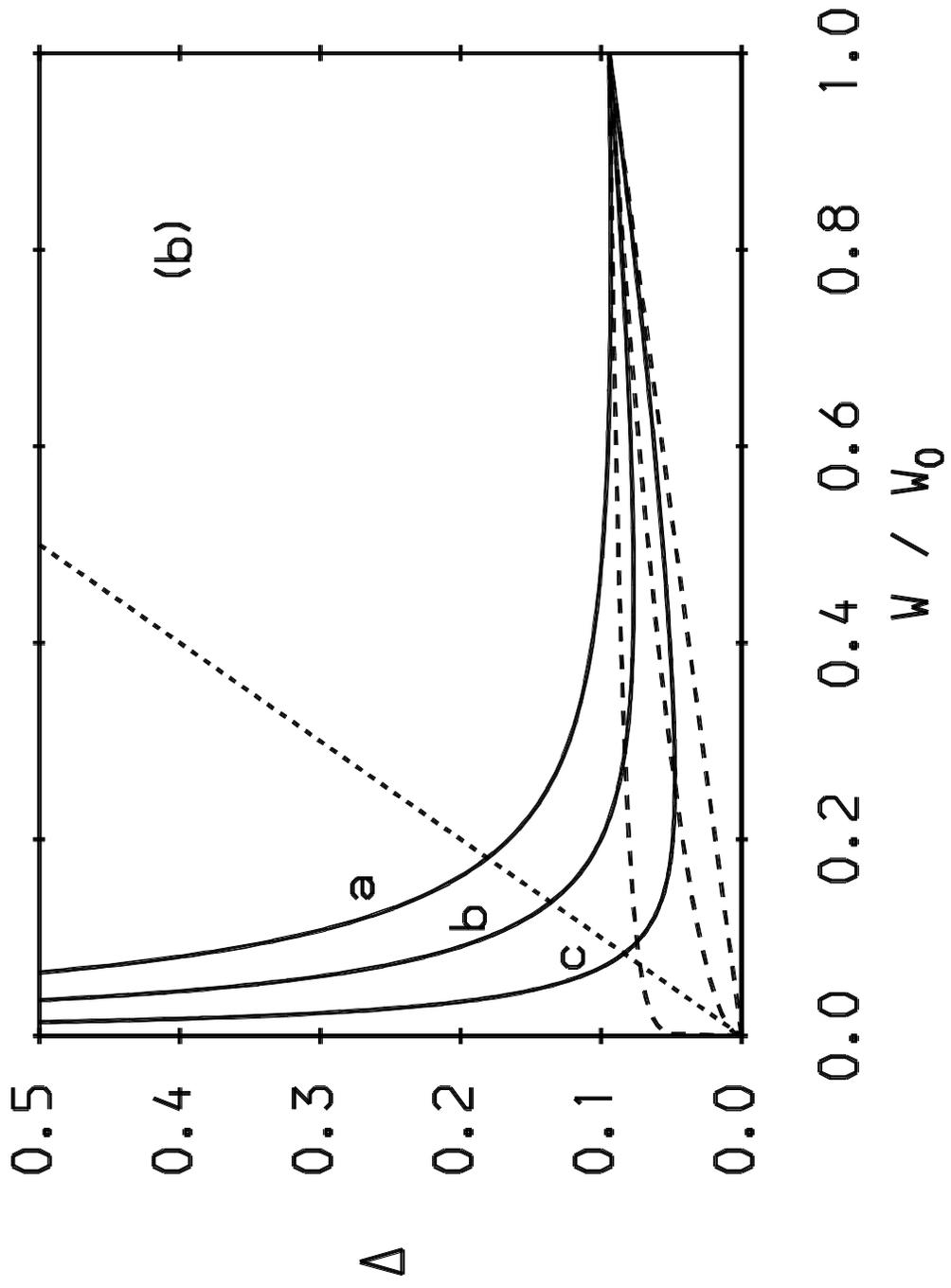
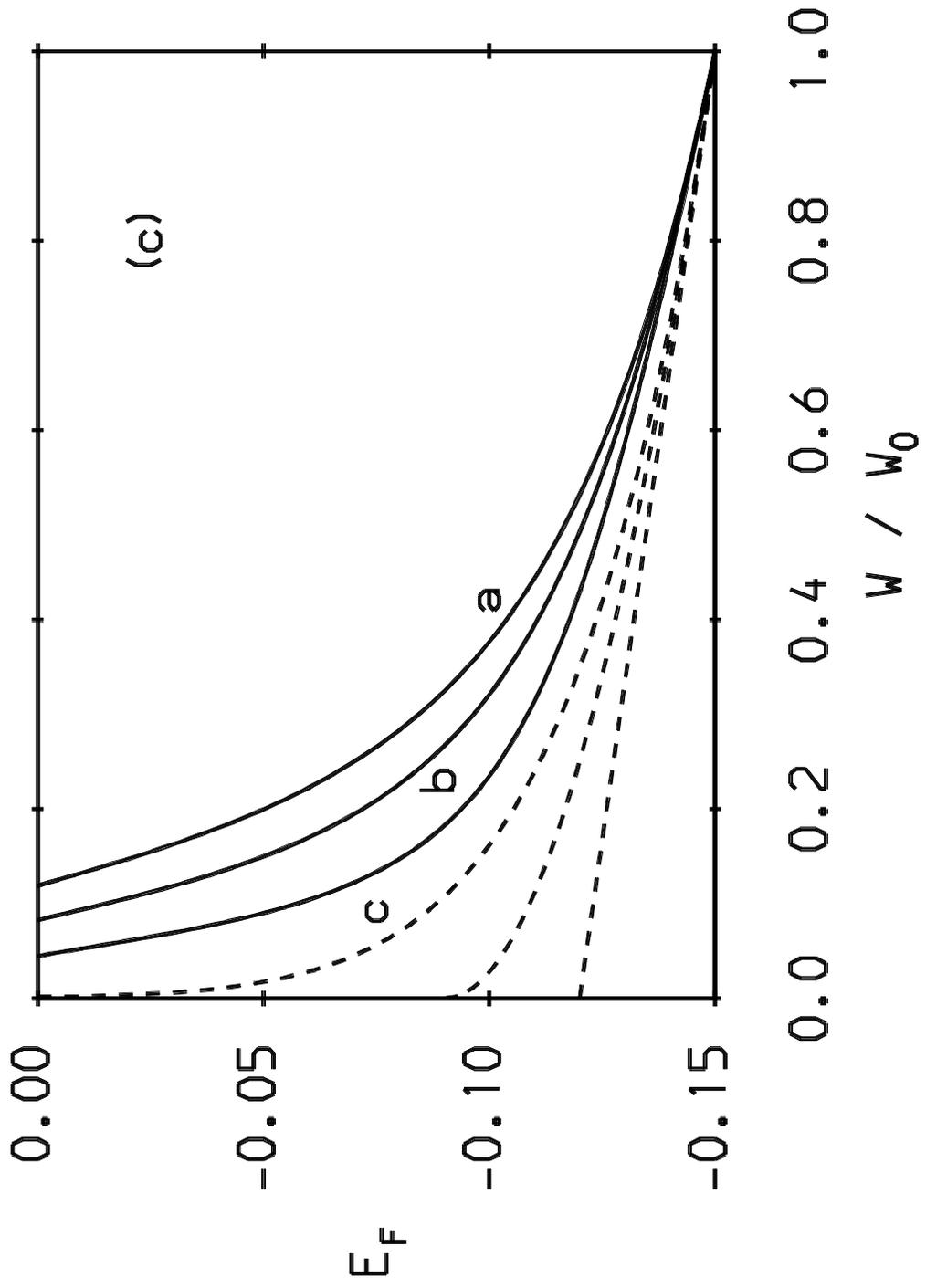
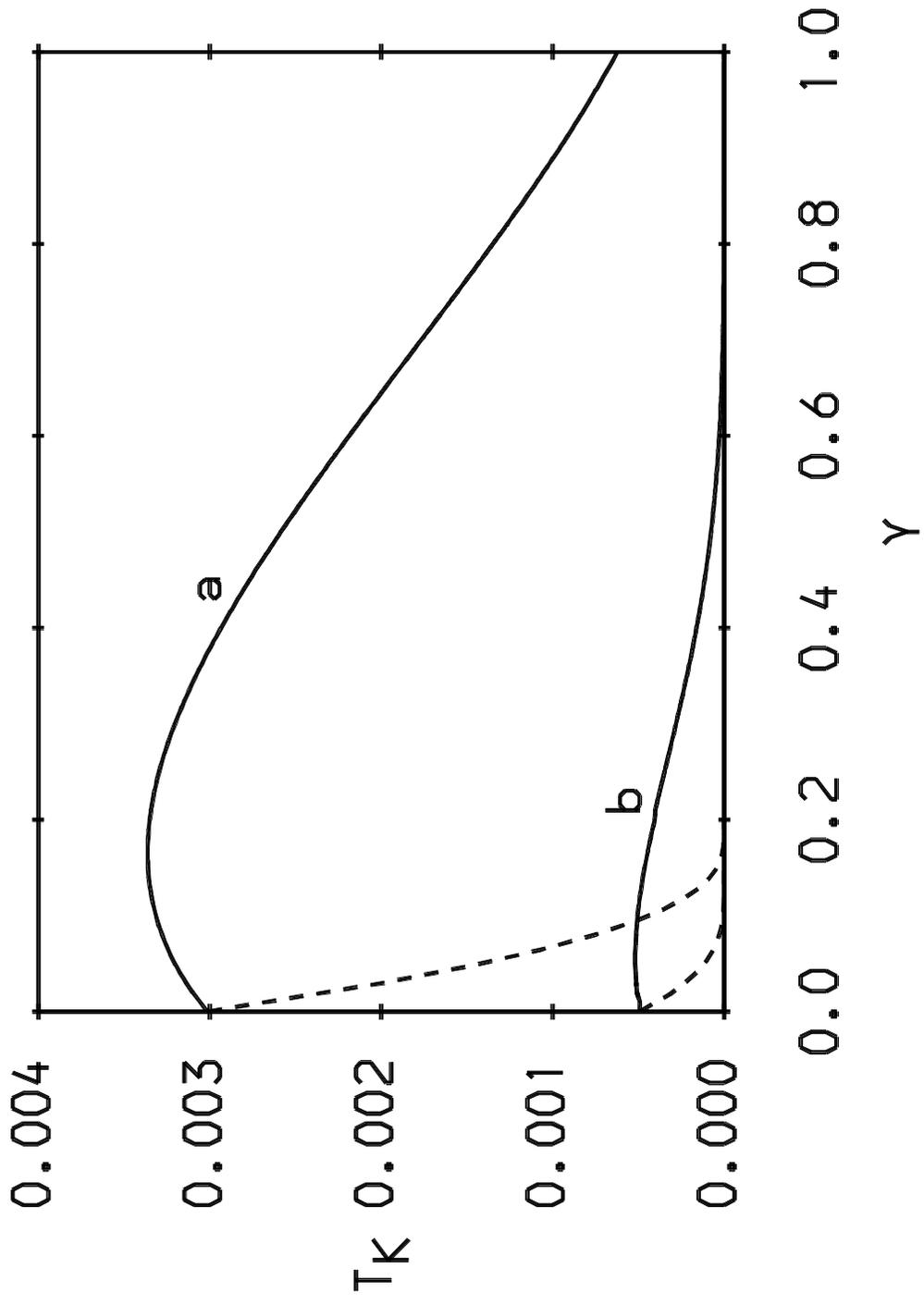
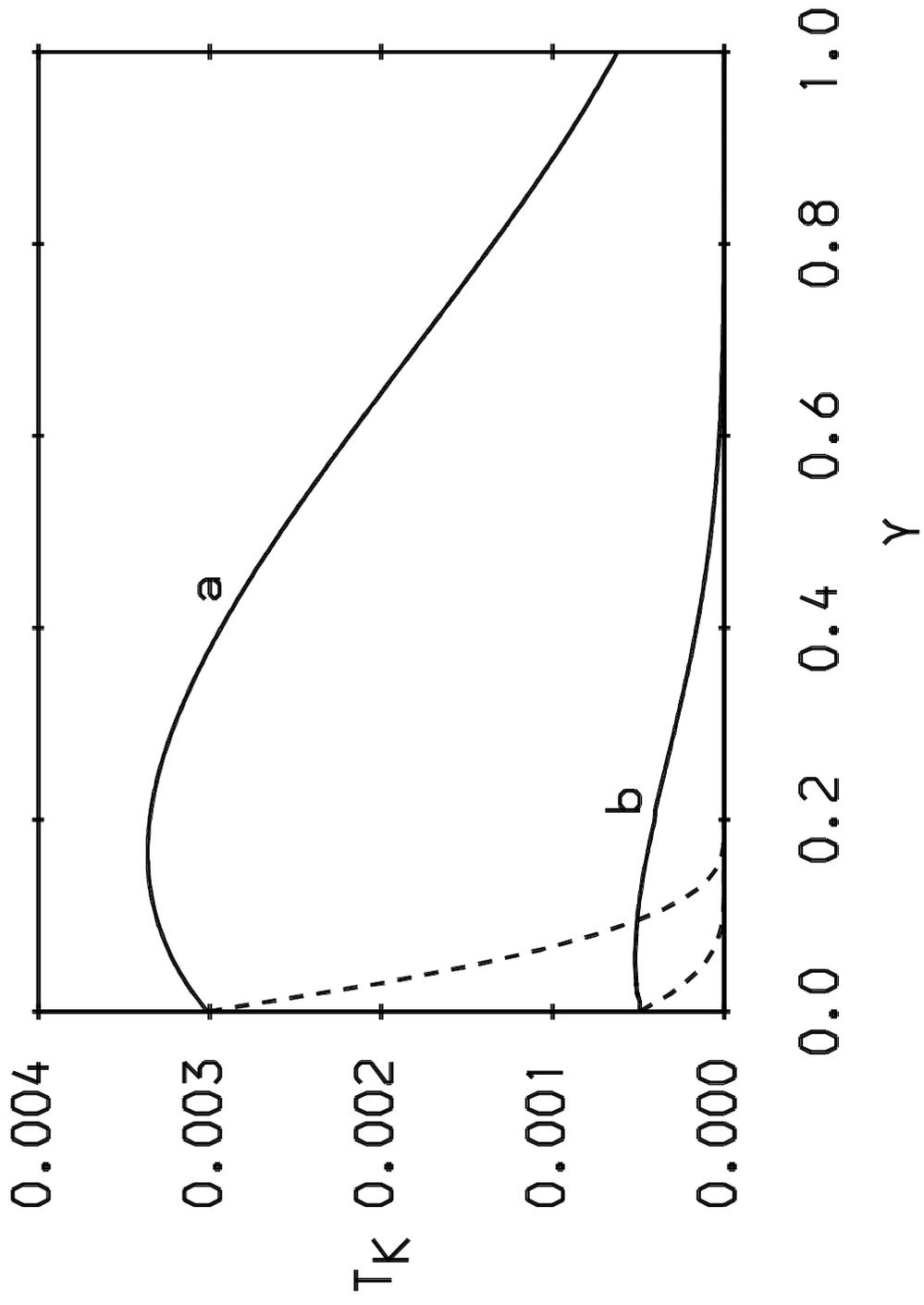
\end{document}